\documentclass[graybox,envcountchap,sectrefs]{svmono}

\usepackage[bookmarks=true,colorlinks=true,linktocpage=true,backref=true]{hyperref}

\usepackage{mathptmx}
\usepackage{helvet}
\usepackage{courier}
\usepackage{type1cm}         

\usepackage{makeidx}         
\usepackage{graphicx}        
\usepackage{multicol}        
\usepackage[bottom]{footmisc}

\makeindex             

\begin{document}

\newcommand{\ud}{\mathrm{d}}

\author{Christine C\'ordula Dantas}
\title{Technical Notes on Classical Electromagnetism}
\subtitle{With Exercises}
\maketitle

\frontmatter

\tableofcontents

\mainmatter

%
%
%
\chapter{Introduction}
\label{p1_ce} 

\abstract*{Each chapter should be preceded by an abstract (10--15 lines long) that summarizes the content. The abstract will appear \textit{online} at \url{www.SpringerLink.com} and be available with unrestricted access. This allows unregistered users to read the abstract as a teaser for the complete chapter. As a general rule the abstracts will not appear in the printed version of your book unless it is the style of your particular book or that of the series to which your book belongs.
Please use the 'starred' version of the new Springer \texttt{abstract} command for typesetting the text of the online abstracts (cf. source file of this chapter template \texttt{abstract}) and include them with the source files of your manuscript. Use the plain \texttt{abstract} command if the abstract is also to appear in the printed version of the book.}


\section{System of units, the electric current and the electric charge}

In these notes, we use the {International System of Units (SI)\label{SI_concept}}, as described in the BIPM\cite{BIPM}. In this system, the {\sl meter} [m] is used as the unit of length, the {\sl kilogram} [kg] as the unit of mass, and the {\sl second} [s] as the unit of time. 

In electrodynamics, one additional basic unit is necessary: the {\sl ampere} [A]\footnote{Named in honor of the French physicist {Andr\'e-Marie Amp\`ere (1775-1836)\label{Ampere_person}}.} as the unit of electric current. According to the BIPM\cite{BIPM}, ``[T]he ampere [A] is that constant current which, if maintained in two straight parallel conductors of infinite length, of negligible circular cross-section, and placed 1 metre apart in vacuum, would produce between these conductors a force equal to $2 \times 10^{–7}$ newton [N] per metre of length.''

The {\sl coulomb} [C]\footnote{Named in honor of the French physicist {Charles-Augustin de Coulomb (1736-1806)\label{Coulomb_person}}.} is a derived unit of electric charge in the SI. One coulomb is the quantity of charge transported by a constant current of 1 [A] in 1 [s]:

\begin{equation}
1 ~{\rm [C]} = 1 ~{\rm [A]} \times 1 ~{\rm [s]}. \label{coulomb_eqn}
\end{equation}

\noindent The most recent value of the {\sl elementary charge}, $e$, is $e = 1.602~176~565(35) \times 10^{-19}$ [C]\cite{CODATA}. The charge of the electron, $e^{-}$, is $e^{-}\equiv-e$, and that of the proton, $e^{+}$, is $e^{+}\equiv +e$.,

In order to have an idea of what do such units and numbers mean in a concrete example, we still need to state a few more concepts. First, the {\sl voltage}, or {\sl electromotive force}, is the potential difference between two points in an electrical field (to be defined in a moment), that is, the energy per unit charge. The unit for the voltage, the {\sl volt} [V]\footnote{Named in honor of the Italian physicist Alessandro Volta (1745-1827).}, is given by:

\begin{equation}
 1 ~~{\rm [V]} = 1 ~~{{\rm [J]}\over {\rm [C]}} = 1 ~~{{\rm [N][m]}\over {\rm [C]}} = 
1 ~~ { {\rm [kg][m]}^2 \over \rm{[C][s]}^2}, \label{volt_eqn}
\end{equation}

\noindent where the {\sl joule} [J]\footnote{Named in honor of the English physicist James Prescott Joule (1818-1889).} is a derived unit of energy or work in the SI, and the {\sl newton} [N]\footnote{Named in honor of the English natural philosopher and mathematician Sir Isaac Newton (1642-1727).} is the SI derived unit of force. Second, the {\sl electric power} (P), that is, the energy output per second, is given by the voltage times the current:

\begin{equation}
P = V \times I  ~~~\left ( {{\rm [J]} \over {\rm [C]}} \times {{\rm [C]} \over {\rm [s]}} \right )= V I ~~ {{\rm [J]} \over {\rm [s]}} = V I ~~{\rm [W]},
\end{equation}

\noindent where the {\sl watt} [W]\footnote{Named in honor of the Scottish engineer James Watt (1736-1819).}, is a derived unit of power in the SI.

Now, for our concrete example, suppose an electric circuit operating on an AC voltage of about $V =120$ [V] (volts). Therefore, in order to use power at the rate of, say $P = 120$ [W] on a $120$ [V] electric circuit would require an electric current of $I = 1$ [A]. Hence, 1 [C] of charge is transported through a $120$ [W] lightbulb in one second in that electric circuit. 


\section{The Coulomb's force law, the Lorentz force law, fields}

Coulomb studied the behavior of two point charged particles, $q_1$ and $q_2$, of equal as well as of opposite charges, and described in 1785 that the force $F_{12}$ between them obeyed an inverse-square law\cite{Cou1785}. Suppose that charge $q_1$ is at the origin of the system of coordinates and $q_2$ is at a distance $r$. Then the force on $q_2$ due to $q_1$ is:

\begin{equation}
\vec{\mathbf{F}}_{12} = {q_1q_2 \over 4 \pi \epsilon_0 r^3} \vec{\mathbf{r}} =
{q_1q_2 \over 4 \pi \epsilon_0 r^2} \hat{\mathbf{r}}, \label{Coulomb_eqn}
\end{equation}

\noindent where $\hat{\mathbf{r}}$ is a unit vector in the direction that joins both particles. Notice the well-known result that charges with opposite signs attract each other, $\vec{\mathbf{F}}_{12} (q_{1:+};q_{2:-})$ or $\vec{\mathbf{F}}_{12} (q_{1:-};q_{2:+}) $, and those with same signs repel each other, $\vec{\mathbf{F}}_{12} (q_{1:+};q_{2:+})$ or $\vec{\mathbf{F}}_{12} (q_{1:-};q_{2:-})$, along the radius vector between them. Note also that the magnitude of the force vector depends directly on the product of the values of the charges and inversely on their square distance.

{\sl Coulomb's electrostatic force law}, as given by Eq. \ref{Coulomb_eqn}, implicitly states that the force between charged particles is transmitted {\sl instantaneously} (namely, the force is transmitted as a so-called {\sl action at a distance}).  On the other hand, the English scientist {Michael Faraday\label{Faraday_person}} (1791-1867) envisioned that the space between charged particles was occupied by objects, the so-called {\sl fields}, which were responsible for mediating the forces between charges, propagating such forces from volume element to volume element until reaching the charges in question. This mechanism opens the route for describing a {\sl finite velocity of propagation} in the formalism. For the moment, however, we will not assume a finite velocity of propagation and will only address this point later in our exposition.


The field concept can be stated as follows. At a given instant, a test charge is placed at any given point of space and it experiences a resulting force, originated from other charges in the region. One could envision the placement, at a given instant of time, of such a test charge in every point of space, and then the measurement of the resulting force on the test particle in question. This procedure is mathematically expressed by a vector function of space and time, called {\sl the electric field intensity}, $\vec{\mathbf{E}}_0(\vec{\mathbf{r}},t)$, such that its dependence at every spatial point and time is given by the electric force per unit charge at that point:

\begin{equation}
\vec{\mathbf{E}}_0(\vec{\mathbf{r}},t) = {\vec{\mathbf{F}}(\vec{\mathbf{r}},t) \over q}. \label{E_eqn}
\end{equation}

\noindent Notice that a field is defined at any given point $\vec{\mathbf{r}}$ even if there is no charge (namely, no {\sl sources}) placed there. 

What happens in the case which the particle is moving with a constant velocity? It is experimentally verified that if a test particle in vacuum is moving with velocity $\vec{\mathbf{v}}$, the resulting force has two components: one independent of $\vec{\mathbf{v}}$ and the other proportional to $\vec{\mathbf{v}}$, but orthogonal to it. This behavior is summarized by the {\sl Lorentz force law}\footnote{Named in honor of the Dutch physicist Hendrik Antoon Lorentz (1853-1928).}, :

\begin{equation}
\vec{\mathbf{F}}=  q \left ( \vec{\mathbf{E}}_0 + \vec{\mathbf{v}} \times \vec{\mathbf{B}}_0  \right ), \label{Lorentz_eqn}
\end{equation}
\noindent where 
\begin{equation}
\vec{\mathbf{B}}_0 \equiv \mu_0  \vec{\mathbf{H}}_0 \label{B_0_eqn}
\end{equation}

\noindent  is the so-called {\sl magnetic flux density} in vacuum and $\vec{\mathbf{H}}_0$ is the {\sl magnetic field intensity}. The proportionality constant $\mu_0$ relating $\vec{\mathbf{B}}_0$ and $\vec{\mathbf{H}}_0$ is called the {\sl vacuum permeability}. Notice that these physical objects are also fields, that is, they may depend on position and time. We will address these quantities in a moment.


\section{The units for the electromagnetic fields}

The Lorentz force law defines the units for $\vec{\mathbf{E}}_0$ and $\vec{\mathbf{B}}_0$. According to Eqs. \ref{E_eqn} and \ref{volt_eqn},

\begin{equation}
1 ~~{\rm [}\vec{\mathbf{E}}_0{\rm ]} = 1 ~~{{\rm [N]} \over {\rm [C]} } = 1~~{{\rm [V]} \over {\rm [m]} }. \label{E_unit_eqn}
\end{equation}

\noindent Also, from Eqs. \ref{Lorentz_eqn} and \ref{E_unit_eqn},

\begin{equation}
1 ~~ {\rm [}\vec{\mathbf{B}}_0{\rm ]}= 
1 {{\rm [N]} \over {\rm [C]} }
{1 \over {\rm [}\vec{\mathbf{v}} {\rm ]} } 
=  1 {{\rm [V][s]} \over {\rm[m]}^2 }\equiv 1 {\rm [T]}, \label{B_0_unit_eqn}
\end{equation}

\noindent where {\sl tesla} [T]\footnote{Named in honor of the Serbian physicist and inventor Nikola Tesla (1856-1943).} is the derived SI unit for the magnetic flux density. 

At this point we cannot know the units for $\vec{\mathbf{H}}_0$ and $\mu_0$ separately. Hence we simply state now the unit for $\vec{\mathbf{H}}_0$ but we will derive it appropriately later, which is:

\begin{equation}
1 ~~{\rm [}\vec{\mathbf{H}}_0{\rm ]} = 1 ~~{{\rm [A]} \over {\rm [m]} }. \label{H_unit_eqn}
\end{equation}

\noindent Therefore, from Eqs. \ref{B_0_eqn}, \ref{B_0_unit_eqn},  \ref{H_unit_eqn},

\begin{equation}
{\rm [}\mu_0{\rm]} = {{\rm [T]}\over {\rm [A]/[m]}} =
 {{\rm [T][m]}\over {\rm [A]}} =
 {{\rm [T][m]}^2\over {\rm [A][m]}} 
\equiv {{\rm [W]} \over {\rm [A][m]}} \equiv {{\rm [h]}\over {\rm [m]} },
\end{equation}

\noindent where we find the derived SI units of {\sl weber} [W]\footnote{Named in honor of the German physicist Wilhelm Eduard Weber (1804-1891).}, and {\sl henry} [h]\footnote{Named in honor of the American scientist Joseph Henry (1797-1878). Notice that we shall adopt a minuscule h in order not to confuse with the magnetic field intensity.}. The numeric value for the vacuum permeability is

\begin{equation}
\mu_0 \simeq 4 \pi \times 10^{-7} ~~ {\rm [h]/[m]}. \label{mu_0}
\end{equation}

It is important to know how to transform these units to the {\sl gaussian} (or {\sl cgs}) system of units, as it is quite usual in the literature to find them quoted on either systems. For the magnetic flux density, we have the {\sl gauss} unit [G]\footnote{Named in honor of the German mathematician and scientist Johann Carl Friedrich Gauss (1777-1855).}, and for the magnetic field intensity, we have the {\sl oersted} unit [Oe]\footnote{Named in honor of the Danish  physicist Hans Christian Oersted (1777-1851).}. We show the conversion from SI to cgs or vice-versa in Tab. \ref{Conv_tab}.

\begin{center}
\begin{table}[htbp]
\centering
\caption{\label{Conv_tab} Conversion between SI and cgs system of units for magnetic field quantities}
\begin{tabular}{|l|c|c|c|}\hline
 type          & field     & cgs  & SI \\ \hline
           &     &  &  \\ 
flux density   & ${\rm [}\vec{\mathbf{B}}_0{\rm ]}$ & [G]  & $10^{-4}$ [T] \\
           &     &  &  \\ 
intensity    & ${\rm [}\vec{\mathbf{H}}_0{\rm ]}$ & [Oe] & ${10^{3}\over 4 \pi}$ [A]/[m]\\ 
           &     &  &  \\ \hline
\end{tabular}
\end{table}
\end{center}

\section*{Problems}
\addcontentsline{toc}{section}{Problems}
%
\begin{prob}
\label{prob1_p1_ce}(*)
Let the magnitude of the magnetic field intensity at a given point measured to be $\mid \vec{\mathbf{H}}_0\mid= 1900$ [Oe]. What is the value of the corresponding magnetic flux density in the SI unit [T]?
\end{prob}

\begin{prob}
\label{prob2_p2_ce}(*)
(Based on \cite{Hau89}, Ch. 1).

\noindent {\bf a)} Combine Newton's law ($\vec{\mathbf{F}}=m \vec{\mathbf{a}}$) with Lorentz law  (Eq. \ref{Lorentz_eqn}) to describe the motion of an electron in an uniform electric field (you may assume that $\vec{\mathbf{E}}_0 = E_x \hat{\mathbf{x}}$), where magnetic fields are negligible. Assume that the electron has an initial velocity $\vec{\mathbf{v}}(t=t_i) \equiv v_i \hat{\mathbf{x}}$, that is, in the same direction of the electric field.

\noindent {\bf b)} Find the value of the electric field $E_x$ so that,  for $E_x>0$, $\alpha$ seconds after releasing the electron with an initial velocity $v_i = \alpha$ [m][s]$^{-1}$, the electron halts and changes direction. Here, $\alpha$ is a positive real number. (Notice that the electric field must be extremely tiny to meet those requirements). 

\noindent {\bf c)} For the conditions stated in item {\bf b}, plot the position of the electron as a function of time for $E_x>0$ and $E_x<0$, considering $1 \leq v_i \leq 5$ in intervals of $\Delta v_i = 0.5$.

\noindent {\bf d)} Assuming $v_i = 0$ and $E_x<0$, show that the electron velocity is given by
\begin{equation}
v = \sqrt{2e |E_x| l/m}
\end{equation}
\noindent where $l>0$ is the value of the electron position. (What is the velocity for $E_x>0$, also with $v_i = 0$?) Fix $E_x = -10^{-2}$ [N]/[C]. Plot $l \times v$ for $10^{-4} \leq l \leq 10$ [m]. (Notice that even with a relatively low value for the electric field, the electron can reach a very high velocity within a short distance). 
\end{prob}

%
%
%
\chapter{Maxwell's equations}
\label{p1_ce2} 

\abstract*{Each chapter should be preceded by an abstract (10--15 lines long) that summarizes the content. The abstract will appear \textit{online} at \url{www.SpringerLink.com} and be available with unrestricted access. This allows unregistered users to read the abstract as a teaser for the complete chapter. As a general rule the abstracts will not appear in the printed version of your book unless it is the style of your particular book or that of the series to which your book belongs.
Please use the 'starred' version of the new Springer \texttt{abstract} command for typesetting the text of the online abstracts (cf. source file of this chapter template \texttt{abstract}) and include them with the source files of your manuscript. Use the plain \texttt{abstract} command if the abstract is also to appear in the printed version of the book.}


\section{Maxwell's equations: macroscopic field}

We have seen that Lorentz force law \ref{Lorentz_eqn} describes the movement of charges given the electromagnetic field. Now we are interested in the effects of charges and their movement (namely, the ``{\sl sources}'') on the electromagnetic field. The sources of an electromagnetic field are expressed in terms of charges and/or charge density currents. In other words, the electric and magnetic fields are originated from charges and currents. 

Those phenomena are described by the {\sl Maxwell's equations}\footnote{Named in honor of the Scottish theoretical physicist and mathematician {James Clerk Maxwell (1831-1879)\label{Maxwell_person}}.}. They can be formulated as either integral or differential equations. The former can be used to determine the fields for symmetrical charge or current configurations. The latter are more appropriate for general problems, as they are applicable at each point of space.

In these notes we are interested in the differential form of Maxwell equations. However, due to their intuitive formulation and beauty, we begin by stating without proof Maxwell's equations in integral form, and then the differential equations can be derived from well-known theorems of vector calculus. We will only state the basic facts without proof.

\subsection{Maxwell's equations in integral form}

$~$

\vspace{0.5cm}
\begin{svgraybox}
\noindent {\sc Gauss's Integral Law} describes how the electric field intensity is related to its source, the charge density:

\begin{equation}
\oint_{S} \epsilon_0 \vec{\mathbf{E}}_0 \cdot \ud \vec{\mathbf{S}} = \int_{V} \rho \ud {\mathrm V}
\label{Gauss_int_eqn}
\end{equation}
\end{svgraybox}
\vspace{0.5cm}

In the equation above, $\ud \vec{\mathbf{S}}$ is a vector corresponding to  an element of area $\ud S$ ($\ud \vec{\mathbf{S}} \equiv \ud S \hat{\mathbf{n}}$; $ \hat{\mathbf{n}}$ is a unit vector normal to the surface at a given point); $\ud V$ is the corresponding element of volume; and $\rho$ is the charge density (net charge per unit volume). The {\sl vacuum (or free-space) permissivity} $\epsilon_0$ is an empirical constant. Let us verify its unit. First, note that the unit of the right hand side quantity in Eq. \ref{Gauss_int_eqn} is [C], because it is just the electric charge inside the volume $V$, delimited by the closed area $S$. Now, in the left hand side, we have inside the integral the scalar product between two vectors, namely, the {\sl electric displacement flux density in vacuum}, defined by

\begin{equation}
\vec{\mathbf{D}}_0\equiv \epsilon_0 \vec{\mathbf{E}}_0, \label{D_0_eqn}
\end{equation} 

\noindent and the area element vector, $\ud \vec{\mathbf{S}}$. As the unit for the area element is [m]$^2$, that of the electric displacement flux density must be, in order to be dimensionally correct, [$\vec{\mathbf{D}}_0$] = [C]/[m]$^2$. Because the unit for $\vec{\mathbf{E}}_0$ is [V]/[m], then we find that
\begin{equation}
{\rm [}\epsilon_0 {\rm ]} = {{\rm [C]} \over {\rm [V][m]}} = {{\rm [F]} \over {\rm [m]}},
\end{equation} 

\noindent where the derived unit [F] is called the {\sl Farad}. The numeric value of the vacuum permissivity is
\begin{equation}
\epsilon_0  \simeq 
{10^{-9} \over 36 \pi} ~{{\rm [F]} \over {\rm [m]}}. \label{epsilon_0}
\end{equation} 

\noindent Notice that we can form a constant from $\epsilon_0$ (Eq. \ref{epsilon_0}) and $\mu_0$ (Eq. \ref{mu_0}), given by (verify the correctness of the final units):
\begin{equation}
c = { 1 \over \sqrt{\epsilon_0 \mu_0}}\simeq 3 \times 10^8 ~~{\rm [m]/[s]}. \label{light_c}
\end{equation} 

\noindent This constant in units of velocity is simply the velocity of light in vacuum, which is the velocity of any electromagnetic wave in vacuum. This fact can be derived from Maxwell's equations, as we shall see later.

\vspace{0.5cm}
\begin{svgraybox}
\noindent {\sc Amp\`ere's Integral Law} describes how the magnetic field intensity is related to its source, the charge current density $\vec{\mathbf{J}}$, plus an additional term involving the electric displacement flux density ${\mathbf{D}}_0\equiv\epsilon_0 \vec{\mathbf{E}}_0$ (c.f. Eq. \ref{D_0_eqn}):

\begin{equation}
\oint_{C}\vec{\mathbf{H}}_0\cdot \ud \vec{\mathbf{l}} = \int_{S} \vec{\mathbf{J}} \cdot \ud \vec{\mathbf{S}}
+ {\ud \over \ud t} \int_{S} \epsilon_0 \vec{\mathbf{E}}_0 \cdot \ud \vec{\mathbf{S}}.
\label{Ampere_int_eqn}
\end{equation}
\end{svgraybox}
\vspace{0.5cm}

In the equation above, the open surface $S$ is delimited by the closed line, or contour, $C$, and $\ud \vec{\mathbf{l}}$ is the line element of the contour $C$. The second term above, given by the time rate of change of the net electric displacement flux density $\epsilon_0 \vec{\mathbf{E}}_0$ through the surface $S$, is named {\sl displacement current}.

{\sl Remark:} The charge density in Gauss's integral law (Eq. \ref{Gauss_int_eqn}) does not include charges that are part of the structure of matter, namely, which are compensated leading to an internal null charge. The same is valid for the charge current density $ \vec{\mathbf{J}}$ in Amp\`ere's integral law (Eq. \ref{Ampere_int_eqn}).

{\sl Remark:} the magnetic field intensity $\vec{\mathbf{H}}_0$ given in Amp\`ere's law (Eq. \ref {Ampere_int_eqn}) is not multiplied by the vacuum permeability, $\mu_0$. It becomes evident, therefore, that the units of the magnetic field intensity are [$\vec{\mathbf{H}}_0$] = [C][s]$^{-1}$[m]$^{-1}$=[A]/[m], as mentioned previously.

{\sl Remark:}  Gauss's (Eq. \ref{Gauss_int_eqn}) e Amp\`ere's  (Eq. \ref{Ampere_int_eqn}) laws relate the field to the corresponding sources, and the latter are related by the charge continuity equation (Eq. \ref{Cont_int_eqn}). The next two equations only involve the fields.

\vspace{0.5cm}
\begin{svgraybox}
\noindent {\sc Faraday's Integral Law} establishes that the circulation of $\vec{\mathbf{E}}_0$ along the contour $C$ is determined by the time rate of change of the magnetic flux density $\vec{\mathbf{B}}_0 \equiv \mu_0 \vec{\mathbf{H}}_0$ (c.f. Eq \ref{B_0_eqn}) through the surface delimitd by the contour:

\begin{equation}
\oint_{C}\vec{\mathbf{E}}_0\cdot \ud \vec{\mathbf{l} }= - {\ud \over \ud t}
\int_{S} \mu_0 \vec{\mathbf{H}} \cdot \ud \vec{\mathbf{S}}.
\label{Faraday_int_eqn}
\end{equation}
\end{svgraybox}
\vspace{0.5cm}

\vspace{0.5cm}
\begin{svgraybox}
\noindent {\sc Gauss's Integral Law for the Magnetic Flux Density} establishes that the net magnetic flux density in any region delimited by a closed surface is zero:

\begin{equation}
\oint_{S} \mu_0 \vec{\mathbf{H}}_0 \cdot \ud \vec{\mathbf{S}} = 0.
\label{Gauss_mag_int_eqn}
\end{equation}
\end{svgraybox}
\vspace{0.5cm}

\subsection{Maxwell's equations in differential form}

$~$

In order to obtain Maxwell's equations in differential form in vacuum, one applies two theorems of calculus on the corresponding integral equations. The relevant theorems are:

\vspace{0.5cm}
\begin{svgraybox}
\noindent {\sc Gauss's integral theorem:} establishes the correspondence between the integral of a vector field $\vec{\mathbf{F}}$ on a closed arbitrary area and the volume integral (delimited by the same area) of the divergent of the same field:
\vspace{0.5cm}
\begin{equation}
\oint_{S}\vec{\mathbf{F}} \cdot \ud \vec{\mathbf{S}} = \int_{V} \nabla \cdot \vec{\mathbf{F}} ~\ud {\textrm V}.
\label{Gauss_teo_eqn}
\end{equation}
\end{svgraybox}
\vspace{0.5cm}

\vspace{0.5cm}
\begin{svgraybox}
\noindent {\sc Stokes integral theorem:} establishes the correspondence between the integral of a closed arbitrary contour of a vector field $\vec{\mathbf{F}}$ and the surface integral (delimited by the same contour) of the rotational of the same field:
\vspace{0.5cm}
\begin{equation}
\oint_{C}\vec{\mathbf{F}} \cdot \ud \vec{\mathbf{l}} = \int_{S} \nabla \times \vec{\mathbf{F}} 
\cdot \ud \vec{\mathbf{S}}.
\label{Stokes_teo_eqn}
\end{equation}
\end{svgraybox}
\vspace{0.5cm}

Maxwell's equations in differential form are now stated without proof. They can be obtained directly from the theorems above (Eqs. \ref{Gauss_teo_eqn} and \ref{Stokes_teo_eqn}) and their derivation are left as an exercise (see Problem \ref{prob6_p1_ce}).

\vspace{0.5cm}
\begin{svgraybox}
\noindent {\sc Gauss's differential law:} 
\vspace{0.5cm}
\begin{eqnarray}
\nabla \cdot \epsilon_0 \vec{\mathbf{E}}_0 & =  & \rho \nonumber\\
                                & {\textrm {or}} & \nonumber\\
\nabla \cdot \vec{\mathbf{D}}_0      & =  & \rho. \label{Gauss_dif_eqn}
\end{eqnarray}
\end{svgraybox}
\vspace{0.5cm}

\begin{svgraybox}
\noindent {\sc Amp\'ere's differential law:} 
\vspace{0.5cm}
\begin{eqnarray}
\nabla \times \vec{\mathbf{H}}_0 & = & \vec{\mathbf{J}} + {\partial \epsilon_0 \vec{\mathbf{E}}_0 \over \partial t}\nonumber\\
                     & {\textrm {or}} & \nonumber\\
\nabla \times \vec{\mathbf{H}}_0 & = & \vec{\mathbf{J}} + {\partial \vec{\mathbf{D}}_0 \over \partial t}.
\label{Ampere_dif_eqn}
\end{eqnarray}
\end{svgraybox}

\vspace{0.5cm}
\begin{svgraybox}
\noindent {\sc Faraday's differential law:} 
\vspace{0.5cm}
\begin{eqnarray}
\nabla \times \vec{\mathbf{E}}_0 & = & - {\partial \mu_0 \vec{\mathbf{H}}_0 \over \partial t} \nonumber\\
                      & {\textrm {or}} & \nonumber\\
\nabla \times \vec{\mathbf{E}}_0 & = & - {\partial \vec{\mathbf{B}}_0 \over \partial t}.
\label{Faraday_dif_eqn}
\end{eqnarray}
\end{svgraybox}
\vspace{0.5cm}

\begin{svgraybox}
\noindent {\sc Gauss's differential law for the magnetic field:} 
\vspace{0.5cm}
\begin{eqnarray}
\nabla \cdot \mu_0 \vec{\mathbf{H}}_0 & = & 0 \nonumber\\
                           & {\textrm {or}} & \nonumber\\
\nabla \cdot \vec{\mathbf{B}}_0 & = & 0.
\label{Gauss_mag_dif_eqn}
\end{eqnarray}
\end{svgraybox}

We also include the charge continuity equation (Eq. \ref{Cont_int_eqn}) in differential form:

\begin{equation}
\nabla \cdot \vec{\mathbf{J}} = - {\partial \rho \over \partial t}.
\label{Cont_dif_eqn}
\end{equation}

\subsection{Continuity conditions}

$~$

It is also important to address how the fields behave in singular configurations. The behavior of the fields under such conditions, for instance, by passing from one side of a charged surface to the other side, is ruled by boundary conditions. The latter can be deduced directly from Maxwell's equations. Here we shall only state those conditions (see Problem \ref{prob7_p1_ce}). 

We denote $\sigma_s$ the surface charge density, and $\vec{\mathbf{K}}$, the surface current density, given respectively by:

\begin{equation}
\sigma_s = \lim_{h \rightarrow 0; ~ \rho \rightarrow \infty} \int_{-h/2}^{+h/2} \rho ~ \ud \xi,
\end{equation}
and
\begin{equation}
\vec{\mathbf{K}} = \lim_{h \rightarrow 0; ~ |\vec{\mathbf{J}}| \rightarrow \infty} \int_{-h/2}^{+h/2} \vec{\mathbf{J}} ~ \ud \xi,
\end{equation}

\noindent where $\xi$ is a coordinate parallel to the normal of the surface, where we first suppose the existence of a layer with thickness $h$, and then shrink it in order to define the singular surface (charge and current) densities accordingly in the limit where the thinkness go to zero. 

The continuity conditions therefore are:

{\sc I:} The normal component of the electric displacement flux density ($\vec{\mathbf{D}}_0 \equiv \epsilon_0 \vec{\mathbf{E}}_0$) is discontinuous in the presence of a surface charge density that separates two regions ($\alpha$, $\beta$):

\begin{equation}
\hat{\mathbf{n}} \cdot \left ( \epsilon_0 \vec{\mathbf{E}}_{0,\alpha} - \epsilon_0 \vec{\mathbf{E}}_{0,\beta} \right ) = \sigma_s.
\label{gauss_cont}
\end{equation}

{\sc II:} The tangential component of the magnetic field intensity $\vec{\mathbf{H}}_0$ is discontinuous in the presence of a surface current density that separates two regions ($\alpha$, $\beta$):

\begin{equation}
\hat{\mathbf{n}} \times\left ( \vec{\mathbf{H}}_{0,\alpha} - \vec{\mathbf{H}}_{0,\beta} \right ) = \vec{\mathbf{K}}.
\label{ampere_cont}
\end{equation}

{\sc III:} The tangential component of the electric field intensity $\vec{\mathbf{E}}_0$ is continuous in the presence of a surface charge density that separates two regions ($\alpha$, $\beta$): 

\begin{equation}
\hat{\mathbf{n}} \times\left ( \vec{\mathbf{E}}_{0,\alpha} - \vec{\mathbf{E}}_{0,\beta} \right ) = 0.
\label{faraday_cont}
\end{equation}

{\sc VI:} The normal component of the magnetic flux density $\vec{\mathbf{B}}_0 \equiv  \mu_0 \vec{\mathbf{H}}_0$ is continuous in the presence of a surface current density that separates two regions ($\alpha$, $\beta$):

\begin{equation}
\hat{\mathbf{n}} \cdot \left ( \mu_0 \vec{\mathbf{H}}_{0,\alpha} - \mu_0 \vec{\mathbf{H}}_{0,\beta} \right ) = 0.
\label{gauss_mag_cont}
\end{equation}

{\sc V:} The normal component of the charge current density is discontinuous between two regions ($\alpha$, $\beta$) only if the surface charge density that separates these regions changes with time:

\begin{equation}
\hat{\mathbf{n}} \cdot \left ( \vec{\mathbf{J}}_{\alpha} - \vec{\mathbf{J}}_{\beta} \right ) = 
- {\partial \sigma_s \over \partial t}.
\label{charge_cons_cont}
\end{equation}

\section{Maxwell's equations: microscopic field}

\subsection{A note on constitutive relations in material media}

$~$

In the previous sections, we have addressed Maxwell's equations in integral and differential forms. The expressions involved the relationship between fields and their sources. We have also noted that the sources appearing in those equations {\sl do not} involve bound charges and current densities {\sl intrinsic} to the matter structure (bound to atoms and molecules). 

Here we remark that the electric displacement flux density, $\vec{\mathbf{D}}_0\equiv \epsilon_0 \vec{\mathbf{E}}_0$ (Eq. \ref{D_0_eqn}), and the magnetic field intensity, $\vec{\mathbf{H}}_0\equiv {\vec{\mathbf{B}}_0 \over \mu_0}$ (Eq. \ref{B_0_eqn}), are regarded as {\sl auxiliary fields}, whereas the electric field intensity,  $\vec{\mathbf{E}}_0$, and the magnetic flux density,  $\vec{\mathbf{B}}_0$, are regarded as {\sl fundamental fields}, as it is clear from Faraday's law (c.f. Eqs. \ref{Faraday_int_eqn} or \ref{Faraday_dif_eqn}) and Gauss's law for the magnetic field (c.f. Eqs. \ref{Gauss_mag_int_eqn} or \ref{Gauss_mag_dif_eqn}) that those fields can exist even is source-free space. 

We also remind that, up to now, we have treated the fields {\sl in vacuum}. In our notation, we have been using the lower index ``$0$'' for the fields to remind us of that fact.  In the presence of matter, we must include the response of the matter itself to the fields when those are applied. The auxiliary fields are related to the fundamental fields via the so-called {\sl constitutive relations}. They can be stated in similar form to those in vacuum, namely:

\begin{equation}
\vec{\mathbf{D}}\equiv \epsilon \vec{\mathbf{E}}, \label{D_const_eqn}
\end{equation}

\begin{equation}
\vec{\mathbf{H}} \equiv {\vec{\mathbf{B}} \over \mu }, \label{H_const_eqn}
\end{equation}

\noindent where now we drop the lower ``$0$''index, and the quantities $\epsilon$ and $\mu$ (the electric permittivity and magnetic permeability, respectively) may assume different forms, for instance: as a constant number (giving a simple linear relation between the fundamental and auxiliary fields); as a second rank tensor (in anisotropic materials); or even as functions of the fundamental fields $\vec{\mathbf{E}}$ and $\vec{\mathbf{B}}$ themselves in case of very strong fields. 

On the other hand, the response of the matter is expressed in the definition of the auxiliary fields by adding to the vacuum field the so-called {\sl electric polarization} field, $\vec{\mathbf{P}}$, and the {\sl magnetization} field, $\vec{\mathbf{M}}$, namely:

\begin{equation}
\vec{\mathbf{D}} \equiv  \epsilon_0\vec{\mathbf{E}} + \vec{\mathbf{P}} = \vec{\mathbf{D}}_0  + \vec{\mathbf{P}}. \label{D_const_eqn_2}
\end{equation}

\begin{equation}
\vec{\mathbf{B}} \equiv  \mu_0\vec{\mathbf{H}}+ \mu_0\vec{\mathbf{M}} = \vec{\mathbf{B}}_0 + \mu_0\vec{\mathbf{M}} \label{H_const_eqn_2}
\end{equation}

If the material is electrically linear and isotropic, one can write a linear relation between $\vec{\mathbf{P}}$ and $\vec{\mathbf{E}}$:

\begin{equation}
\vec{\mathbf{P}} =  \chi_e \epsilon_0 \vec{\mathbf{E}}, \label{chi_e}
\end{equation}

\noindent where $\chi_e$ is the  {\sl dielectric susceptibility}. It is clear from Eqs. \ref{D_const_eqn}, \ref{D_const_eqn_2} and \ref{chi_e} that the material electric permittivity is then:

\begin{equation}
\epsilon \equiv \epsilon_0(1 + \chi_e).
\end{equation}

Similarly, for linearly magnetized materials:

\begin{equation}
\vec{\mathbf{M}} = \chi_m \vec{\mathbf{H}}, \label{chi_m}
\end{equation}

\noindent where $\chi_m$ is the  {\sl magnetic susceptibility}, giving, from Eqs. \ref{H_const_eqn}, \ref{H_const_eqn_2} and \ref{chi_m}, the material magnetic permeability:

\begin{equation}
\mu \equiv \mu_0(1 + \chi_m).
\end{equation}

{\sl Remark:} The nature of the polarization and magnetization fields is the following. We define the total charge density as a two-component object, given by:

\begin{equation}
\rho = \rho_{\rm free} + \rho_{\rm pair},
\end{equation}

\noindent where the first term refers to charges that can move freely from an atomic site to the other, as in a conductor. The second term refers to paired charges in which an electric field induces {\sl electric dipole moments},

\begin{equation}
\vec{\mathbf{p}} = q \vec{\mathbf{d}},
\end{equation}

\noindent where $\vec{\mathbf{d}}$ is the distance between a pair of opposite charges. Then, the polarization vector is defined as 

\begin{equation}
\vec{\mathbf{P}} \equiv N \vec{\mathbf{p}},
\end{equation}

\noindent where $N$ is the number of polarized particles by volume unit. 

Similarly, the sources of magnetic field in material media are the {\sl magnetic dipoles} $\vec{\mathbf{m}}$, which can be approximately aligned, and arise from individual charges or currents caused by ``circularly'' moving charges. Analogously to electric dipoles, we identify:

\begin{equation}
\vec{\mathbf{p}} \leftrightarrow \mu_0 \vec{\mathbf{m}},
\end{equation}

\noindent and, therefore:

\begin{equation}
\vec{\mathbf{M}} = N \vec{\mathbf{m}}.
\end{equation}

{\sl Remark:} It can be shown that:

\begin{equation}
\rho_{\rm pair} = - \nabla \cdot \vec{\mathbf{P}}. \label{rho_pair}
\end{equation}

\noindent and also the ``{\sl magnetic charge density}'' can be defined as:

\begin{equation}
\rho_{\rm m} \equiv - \nabla  \cdot \mu_0 \vec{\mathbf{M}}. \label{rho_m}
\end{equation}

\noindent Inserting Eqs. \ref{rho_pair} and \ref{rho_m} into the divergence of Eqs. \ref{D_const_eqn_2} and \ref{H_const_eqn_2}, respectively, it is clear that:

\begin{eqnarray}
\nabla \cdot \vec{\mathbf{D}} & = & \nabla \cdot \epsilon_0\vec{\mathbf{E}} + \nabla \cdot \vec{\mathbf{P}} \nonumber \\
                     & = & \rho_{\rm total} - \rho_{\rm pair} \nonumber\\
                     & = & \rho_{\rm free}, \label{Gauss_dif_mat}
\end{eqnarray}

\noindent and

\begin{eqnarray}
\nabla \cdot \vec{\mathbf{B}} & = & \nabla \cdot \mu_0\vec{\mathbf{H}} + \nabla \cdot \mu_0  \vec{\mathbf{M}} \nonumber \\
                     & = & \rho_{\rm m} - \rho_{\rm m} \nonumber\\
                     & = & 0. \label{Gauss_dif_mag_mat}
\end{eqnarray}

We list below the general form of {\sl Maxwell's differential equations in material media}:

\begin{equation}
\nabla \cdot \vec{\mathbf{D}} = \rho_{\rm free},
\end{equation}

\begin{equation}
\nabla \cdot \vec{\mathbf{B}}= 0,
\end{equation}

\begin{equation}
\nabla \times \vec{\mathbf{E}} = - {\partial \vec{\mathbf{B}} \over \partial t},
\end{equation}

\begin{equation}
\nabla \times \vec{\mathbf{H}} = \vec{\mathbf{J}}_{\rm free} + {\partial \vec{\mathbf{D}} \over \partial t}.
\end{equation}

\subsection{Microscopic fields}

$~$

In the previous sections, we addressed Maxwell's equations treating the sources in a {\sl macroscopic} view. In other words, the charge $\rho_{\rm free}$ and the current density $\vec{\mathbf{J}}_{\rm free}$ are assumed to be continuous functions of position. In order to meet such a condition, they must be treated, respectively, as an average charge and average current, defined as the charge $\Delta Q$ in a unit volume $\Delta V$, and the current $\Delta I$ across a surface area $\Delta A$, where $\Delta V$ and $\Delta A$ are much larger than atomic or molecular scales. Correspondingly, the electric and magnetic fields are local spatial averages of the {\sl microscopic} fields, the latter given by each charge contribution. 

In order to obtain the microscopic view of Maxwell's equations, we re-define the charge density and the current density in order to address the discreteness of their distribution, by the use of the Dirac delta function, respectively:

\begin{equation}
\rho \equiv \rho(\vec{\mathbf{r}}) = \sum_{\alpha} e_{\alpha} \delta(\vec{\mathbf{r}}-\vec{\mathbf{q}}_{\alpha} ), \label{rho_discrete}
\end{equation}
\noindent and
\begin{equation}
\vec{\mathbf{j}} \equiv \vec{\mathbf{j}}(\vec{\mathbf{r}}) = \sum_{\alpha} e_{\alpha} \dot{\vec{\mathbf{q}}}_{\alpha}\delta(\vec{\mathbf{r}}-\vec{\mathbf{q}}_{\alpha} ), \label {j_discrete}
\end{equation}

\noindent where $\alpha$ labels each charge $e_{\alpha}$, located at coordinate $\vec{\mathbf{q}}_{\alpha}$ and with velocity $\dot{\vec{\mathbf{q}}}_{\alpha}$. We use lower case letters to designate all the microscopic vector fields. Notice that, since all charge particles now contribute to the charge density and to the current density, as it is clear from Eqs. \ref{rho_discrete} and \ref{j_discrete}, we {\sl no longer make reference to the auxiliary fields}, $\vec{\mathbf{D}}$ (Eqs. \ref{D_const_eqn}, \ref{D_const_eqn_2}) and $\vec{\mathbf{H}}$ (Eqs. \ref{H_const_eqn}, \ref{H_const_eqn_2}), as well as to the electric permittivity and magnetic permeability in material media. We then write the {\sl microscopic Maxwell equations in differential form} as:

\begin{equation}
\nabla \cdot \vec{\mathbf{e}} = {\rho \over \epsilon_0}, \label{Gauss_mic}
\end{equation}

\begin{equation}
\nabla \cdot \vec{\mathbf{b}} = 0, \label{Gauss_b_mic}
\end{equation}

\begin{equation}
\nabla \times \vec{\mathbf{e}} = -{\partial \vec{\mathbf{b}} \over \partial t} \label{Faraday_mic}
\end{equation}

\begin{equation}
\nabla \times \vec{\mathbf{b}} = 
\mu_0 \vec{\mathbf{j}} +
\epsilon_0 \mu_0 {\partial \vec{\mathbf{e}} \over \partial t}.
\end{equation}

Given the relation in Eq. \ref{light_c}, Amp\`ere's law will be written as

\begin{equation}
\nabla \times \vec{\mathbf{b}} = 
{1\over \epsilon_0 c^2} \vec{\mathbf{j}} +
{1\over c^2} {\partial \vec{\mathbf{e}} \over \partial t}, \label{Ampere_mic}
\end{equation}

\noindent as we choose to express quantities in terms or $\epsilon_0$ and $c$ only.

\section*{Problems}
\addcontentsline{toc}{section}{Problems}
%

\begin{prob}
\label{prob3_p1_ce}(*)
Derive Coulomb's law (Eq. \ref{Coulomb_eqn}) from Gauss's integral law (Eq. \ref{Gauss_int_eqn}) and Lorentz force law (Eq. \ref{Lorentz_eqn}).
\end{prob}

\begin{prob}
\label{prob4_p1_ce}(*)
Derive the {\sl charge conservation law} (or {\sl continuity law}),
\begin{equation}
\oint_{S} \vec{\mathbf{J}} \cdot \ud \vec{\mathbf{S}} = - {\ud \over \ud t} \int_{V} \rho \ud 
{\mathrm V}, \label{Cont_int_eqn}
\end{equation}
from Gauss's integral law (Eq. \ref{Gauss_int_eqn}) and Amp\`ere's (Eq. \ref{Ampere_int_eqn}).
\end{prob}

\begin{prob}
\label{prob5_p1_ce}(*)
Discuss the physical interpretation of a region where the electric field intensity does not have a circulation. What are the shapes to be chosen for the contour $C$ so that Faraday's law is valid in that case? Show an example.
\end{prob}

\begin{prob}
\label{prob6_p1_ce}
Obtain Maxwell's equations in differential form and the equation of charge continuity from their corresponding integral equations, using Gauss and Stokes theorems (Eqs. \ref{Gauss_teo_eqn} and \ref{Stokes_teo_eqn}).
\end{prob}

\begin{prob}
\label{prob7_p1_ce}
Obtain all the continuity equations of this section. 
\end{prob}

%
%
%
\chapter{Potentials and Gauge Transformations}
\label{p1_ce3} 

\abstract*{Each chapter should be preceded by an abstract (10--15 lines long) that summarizes the content. The abstract will appear \textit{online} at \url{www.SpringerLink.com} and be available with unrestricted access. This allows unregistered users to read the abstract as a teaser for the complete chapter. As a general rule the abstracts will not appear in the printed version of your book unless it is the style of your particular book or that of the series to which your book belongs.
Please use the 'starred' version of the new Springer \texttt{abstract} command for typesetting the text of the online abstracts (cf. source file of this chapter template \texttt{abstract}) and include them with the source files of your manuscript. Use the plain \texttt{abstract} command if the abstract is also to appear in the printed version of the book.}

\section{The electromagnetic scalar and vector potentials}

We have seen in Problem \ref{prob5_p1_ce} that for a static electric field intensity which does not have a circulation, one could define an electromotive force between two points whose calculation does not depend on the chosen path between those points. Note that for a static electric field, Faraday's law (Eq. \ref{Faraday_mic}) requires

\begin{equation}
\nabla \times \vec{\mathbf{e}} = 0.
\end{equation}

By Stoke's theorem (Eq. \ref{Stokes_teo_eqn}), we have

\begin{equation}
\int_{S} \nabla \times \vec{\mathbf{e}} 
\cdot \ud \vec{\mathbf{S}} = \oint_{C}\vec{\mathbf{e}} \cdot \ud \vec{\mathbf{l}} = 0,
\end{equation}

\noindent or

\begin{equation}
\left [ \int_{a}^{b}\vec{\mathbf{e}} \cdot \ud \vec{\mathbf{l}} \right ]_{\rm path~ I} =
\left [ \int_{a}^{b}\vec{\mathbf{e}} \cdot \ud \vec{\mathbf{l}}^{\prime} \right ]_{\rm path~ II},
\end{equation}

\noindent and we see that the static field $\vec{\mathbf{e}}$ is {\sl conservative}. Then we can set up a reference, fixed point, $\vec{\mathbf{r}}_{\rm ref}$, and a given path integral is simply a scalar function $\Phi$ (the {\sl electric potential}) of the final integration point, $\vec{\mathbf{r}}$, namely

\begin{equation}
\Phi(\vec{\mathbf{r}}) - \Phi(\vec{\mathbf{r}}_{\rm ref}) = \int_{\vec{\mathbf{r}}}^{\vec{\mathbf{r}}_{\rm ref}}\vec{\mathbf{e}} \cdot \ud \vec{\mathbf{l}}.
\end{equation}

Before proceeding our analysis, we define: 

\begin{itemize}
\item{
{\sl Solenoidal or Transverse or Divergence-free or} $F^{\perp}$ {\sl fields}:
\begin{equation}
\nabla \cdot \vec{\mathbf{F}}^{\perp} = 0; \label{transverse}
\end{equation}
}
\item{
{\sl Irrotational or Longitudinal or Curl-free or} $F^{\parallel}$ {\sl fields}: 
\begin{equation}
\nabla \times \vec{\mathbf{F}}^{\parallel} = 0. \label{longitudinal}
\end{equation}
}
\end{itemize}

We recall some important results from vector analysis regarding those fields.

\vspace{0.5cm}
\begin{svgraybox}
{\sc Theorem for Solenoidal Fields:} The following conditions are equivalent:
\begin{enumerate}
\item{$\nabla \cdot \vec{\mathbf{F}}^{\perp} = 0$ throughout space.}
\item{$\int_{S}\vec{\mathbf{F}}^{\perp}\cdot\ud \vec{\mathbf{S}}$ does not depend on the surface.}
\item{$\oint_{S}\vec{\mathbf{F}}^{\perp}\cdot\ud \vec{\mathbf{S}}=0$ for any closed surface.}
\item{$\vec{\mathbf{F}}^{\perp} = \nabla \times \vec{\mathbf{A}}$, where $\vec{\mathbf{A}}$ is called the {\sl vectorial potential} of the $\vec{\mathbf{F}}^{\perp}$ field. The field $\vec{\mathbf{A}}$ is not unique, because taking $\vec{\mathbf{A}}^{\prime} \rightarrow \vec{\mathbf{A}}+ \nabla\Phi(\vec{\mathbf{r}})$, we have $\nabla \times \vec{\mathbf{A}} = \nabla \times \vec{\mathbf{A}}^{\prime}$, as $\nabla \times \nabla\Phi(\vec{\mathbf{r}})= 0$.}
\end{enumerate}
\end{svgraybox}

\vspace{1cm}

\begin{svgraybox}
{\sc Theorem for Irrotational Fields:} The following conditions are equivalent:
\begin{enumerate}
\item{$\nabla \times \vec{\mathbf{F}}^{\parallel} = 0$ throughout space.}
\item{$\int_a^b\vec{\mathbf{F}}^{\parallel}\cdot\ud \vec{\mathbf{l}}$ does not depend on the path.}
\item{$\oint_{C}\vec{\mathbf{F}}^{\parallel}\cdot\ud \vec{\mathbf{l}}=0$ for any closed contour.}
\item{$\vec{\mathbf{F}}^{\parallel} = -\nabla \Phi(\vec{\mathbf{r}})$, where $\Phi(\vec{\mathbf{r}})$ is called the  {\sl scalar potential} of the $\vec{\mathbf{F}}^{\parallel}$ field. The field $\Phi(\vec{\mathbf{r}})$ is not unique, because taking $\Phi^{\prime}(\vec{\mathbf{r}}) \rightarrow \Phi(\vec{\mathbf{r}})+ b$, where $b$ is a constant, we have $\nabla \Phi(\vec{\mathbf{r}}) = \nabla \Phi^{\prime}(\vec{\mathbf{r}})$, as $\nabla b= 0$.}
\end{enumerate}
\end{svgraybox}
\vspace{0.5cm}

Given the theorems above, we proceed by applying the theorem for solenoidal fields to the magnetic field $\vec{\mathbf{b}}$ in Eq. \ref{Gauss_b_mic}, so that we find:

\begin{equation}
\vec{\mathbf{b}}^{\perp} = \nabla \times \vec{\mathbf{a}}, \label{vector_a}
\end{equation}

\noindent where $ \vec{\mathbf{a}}$ is the {\sl electromagnetic vector potential}. Note, however, that $\vec{\mathbf{a}}$ cannot be uniquely defined, as we can take $\vec{\mathbf{a}}^{\prime} \rightarrow \vec{\mathbf{a}}+ \nabla\chi$, resulting in the same $\vec{\mathbf{b}}^{\perp}$, because $\nabla \times \nabla\chi= 0$. We call $\chi$ the {\sl gauge function}. 

We proceed by inserting Eq. \ref{vector_a} in Eq. \ref{Faraday_mic}, leading to:

\begin{equation}
\nabla \times \vec{\mathbf{e}} = -{\partial \over \partial t} \left ( \nabla \times \vec{\mathbf{a}} \right )  = - \nabla \times {\partial \vec{\mathbf{a}} \over \partial t} ~ \Longrightarrow ~ \nabla \times
\left ( \vec{\mathbf{e}}  + {\partial \vec{\mathbf{a}} \over \partial t} \right ) = 0 .
\end{equation}

\noindent The equation above implies that, given the theorem for irrotational fields,

\begin{equation}
\left ( \vec{\mathbf{e}}  + {\partial \vec{\mathbf{a}} \over \partial t} \right )^{\parallel}  = -\nabla \phi , \label{scalar_phi}
\end{equation}

\noindent where $\phi$ is the {\sl electromagnetic scalar potential}. We note that the freedom introduced by the gauge function can be expressed by the following transformations, which must be both considered in order that the electromagnetic fields (and Maxwell's equations) remain unaffected:

\begin{eqnarray}
\vec{\mathbf{a}}^{\prime} & \rightarrow & \vec{\mathbf{a}}+ \nabla\chi \nonumber \\
\phi^{\prime}             & \rightarrow & \phi - {\partial \chi \over \partial t} \label{gauge_trans}
\end{eqnarray}

Up to this point, we have analysed the electromagnetic potentials without relating to the sources. By inserting Eq. \ref{scalar_phi} into Eq. \ref{Gauss_mic}, we find:

\begin{equation}
\nabla \cdot \vec{\mathbf{e}}^{\parallel} = 
- \nabla  \cdot {\partial \vec{\mathbf{a}} \over \partial t} - 
\nabla^2 \phi =
- {\partial \over \partial t} \left ( \nabla \cdot \vec{\mathbf{a}} \right )
- \nabla^2 \phi = \rho/ \epsilon_0 , \label{e_parallel}
\end{equation}

\begin{equation}
\Longrightarrow ~\nabla^2 \phi + {\partial \over \partial t} \left ( \nabla \cdot \vec{\mathbf{a}} \right ) = - \rho/ \epsilon_0 , \label{Pot_source_1}
\end{equation}

\noindent and we obtain our first relation connecting the potentials and the electric charge density. Now inserting Eqs. \ref{vector_a} and \ref{scalar_phi} into Eq. \ref{Ampere_mic}, and recalling a well-known vector identity, we find a second relation:

\begin{eqnarray}
\nabla \times \left ( \nabla \times \vec{\mathbf{a}} \right ) & \equiv & -  \nabla^2 \vec{\mathbf{a}} + \nabla
\left ( \nabla \cdot \vec{\mathbf{a}} \right ) \\ \nonumber
 & = & -{1\over c^2} {\partial^2 \vec{\mathbf{a}} \over \partial t^2}
-{1\over c^2} {\partial  \over \partial t} \nabla \phi
+{1\over \epsilon_0 c^2} \vec{\mathbf{j}} 
\end{eqnarray}

\begin{equation}
\Longrightarrow ~ 
\nabla^2 \vec{\mathbf{a}} 
- {1\over c^2} {\partial^2 \vec{\mathbf{a}} \over \partial t^2}
- \nabla \left ( \nabla \cdot \vec{\mathbf{a}} \right )
- {1\over c^2} \nabla \left ( {\partial  \phi \over \partial t}  \right )
= {1\over \epsilon_0 c^2} \vec{\mathbf{j}}. \label{Pot_source_2}
\end{equation}

\section{Gauge transformations: the Lorentz and the Coulomb gauges}

In the previous section, we have derived expressions relating the potentials to the sources, namely, Eqs. \ref{Pot_source_1} and \ref{Pot_source_2}. Notice that those expressions can be simplified because we know that the gauge transformations (Eq. \ref{gauge_trans}) leave the fields invariant, so we may appropriately change the vector and scalar potentials in those equations in order to simplify them without altering the final relations. 

One first choice would be

\begin{equation}
\nabla \cdot \vec{\mathbf{a}}  = - {1 \over c^2} {\partial \phi \over \partial t}. \label{Lorentz_gauge}
\end{equation}

\noindent To show that such a choice can always be made is left as an exercise (\ref{prob8_p1_ce}). We substitute Eq. \ref{Lorentz_gauge} into Eqs. \ref{Pot_source_1} and \ref{Pot_source_2} and obtain, respectively:

\begin{equation}
\left ( \nabla^2 - {1 \over c^2} {\partial^2 \over \partial t^2} \right ) \phi  = - \rho / \epsilon_0,
\end{equation}

\begin{equation}
\left ( \nabla^2 - {1 \over c^2} {\partial^2 \over \partial t^2} \right ) \vec{\mathbf{a}}  = -{1 \over \epsilon_0 c^2}\vec{\mathbf{j}}.
\end{equation}

\noindent We arrive then at a situation where we have two equations {\sl separately relating the scalar potential to the charges and the vector potential to the currents}.

A second choice to simplify Eqs. \ref{Pot_source_1} and \ref{Pot_source_2} by the use of the gauge transformations (Eq. \ref{gauge_trans}) would be the {\sl Coulomb gauge}, given by

\begin{equation}
\nabla \cdot \vec{\mathbf{a}}  = 0. \label{Coulomb_gauge}
\end{equation}

\noindent To show that one can always find such a solenoidal or transverse vector potential is also left as an exercise (\ref{prob9_p1_ce}). Inserting Eq. \ref{Coulomb_gauge} into Eqs. \ref{Pot_source_1} and \ref{Pot_source_2} give us, respectively:

\begin{equation}
\nabla^2 \phi = - \rho/ \epsilon_0 , \label{Pot_source_1_Coul}
\end{equation}

\begin{equation}
\left ( 
\nabla^2 - {1\over c^2} {\partial^2  \over \partial t^2}
\right ) \vec{\mathbf{a}}
= {1\over c^2} \nabla \left ( {\partial  \phi \over \partial t}  \right )
- {1\over \epsilon_0 c^2} \vec{\mathbf{j}}. \label{Pot_source_2_Coul}
\end{equation}

\noindent This time we arrive at a situation where the {\sl scalar potential is related to the instantaneous configuration of charges}, and the {\sl vector potential is related to dynamic sources or currents}.

The choice of the Lorentz gauge or the Coulomb gauge depends on the physical problem. The former is used to problems where the covariant aspect of the theory must be preserved, as in relativistic quantum field theory. The latter is appropriate for slow-moving particles in bound states, as the treatment can be simplified for that choice of gauge. 

From Eq. \ref{vector_a}, we see that $\vec{\mathbf{b}}$ is purely  transverse, and from Eqs. \ref{e_parallel} and \ref{Faraday_mic} (together with Eq. \ref{vector_a}), $\vec{\mathbf{e}}$ can be separated into transversal and parallel components, namely:

\begin{equation}
\vec{\mathbf{b}}^{\perp} = \nabla \times \vec{\mathbf{a}},
\end{equation}

\begin{equation}
\nabla \cdot \vec{\mathbf{e}}^{\parallel} = 
 \rho/ \epsilon_0 , \label{Gauss_parallel}
\end{equation}

\begin{equation}
\nabla \times \vec{\mathbf{e}}^{\perp} = 
 -{\partial \vec{\mathbf{b}}^{\perp} \over \partial t}.
\end{equation}

Also, Amp\`ere's equation gets divided into two components:

\begin{equation}
\nabla \times \vec{\mathbf{b}}^{\perp} = 
{1 \over c^2}{\partial \vec{\mathbf{e}}^{\perp} \over \partial t} +
{1 \over \epsilon_0 c^2}\vec{\mathbf{j}}^{\perp}, 
\end{equation}

\begin{equation}
0 = 
{1 \over c^2}{\partial \vec{\mathbf{e}}^{\parallel} \over \partial t} +
{1 \over \epsilon_0 c^2}\vec{\mathbf{j}}^{\parallel}, 
\end{equation}

\noindent which, by taking the divergerce of the latter, and using Eq. \ref{Gauss_parallel}, we obtain the continuity equation:

\begin{equation}
\nabla \cdot \vec{\mathbf{j}}^{\parallel}+
{\partial \rho \over \partial t} = 0.
\end{equation}

We have seen that the eletric field can be expressed in terms of the time changes in the vector potential and spatial changes in the scalar potential (Eq. \ref{scalar_phi}). For the Coulomb gauge (Eq. \ref{Coulomb_gauge}), $\nabla \cdot \vec{\mathbf{a}}  = 0 \Rightarrow \vec{\mathbf{a}} = \vec{\mathbf{a}} ^{\perp}$ (it is a transverse field, Eq. \ref{transverse}), and we also have necessarily $\nabla \phi = (\nabla \phi)^{\parallel}$ (the gradient is always a longitudinal field, Eq. \ref{longitudinal}). Then, taking those facts into consideration in Eq. \ref{scalar_phi} we find, for the Coulomb gauge:

\begin{equation}
\left ( \vec{\mathbf{e}}^{\parallel} + \vec{\mathbf{e}}^{\perp}   + {\partial \vec{\mathbf{a}}^{\perp} \over \partial t} \right )^{\parallel} = -(\nabla \phi)^{\parallel},
\end{equation}

\begin{equation}
\Rightarrow \vec{\mathbf{e}}^{\parallel} = -(\nabla \phi)^{\parallel},
\end{equation}

\begin{equation}
\Rightarrow \vec{\mathbf{e}}^{\perp}  = - {\partial \vec{\mathbf{a}}^{\perp} \over \partial t}.
\end{equation}

\noindent Therefore, Eq. \ref{Pot_source_2_Coul} can be written in terms of transverse quantities only, giving:

\begin{equation}
\left ( 
\nabla^2 - {1\over c^2} {\partial^2  \over \partial t^2}
\right ) \vec{\mathbf{a}}^{\perp}
= - {1\over \epsilon_0 c^2} \vec{\mathbf{j}}^{\perp}, \label{Pot_source_3_Coul}
\end{equation}

\noindent whereas the static source equation (Eq. \ref{Pot_source_1_Coul}), described by the scalar potential, remains unchanged. Therefore we see that the Coulombic fields (Eq. \ref{Pot_source_1_Coul}) are completely separated from the transverse fields (Eq. \ref{Pot_source_3_Coul}).

Notice that, since $\nabla \chi = (\nabla \chi)^{\parallel}$, gauge transformations (Eq. \ref{gauge_trans}) cannot change $\vec{\mathbf{a}}^{\perp}$, hence it is a {\sl gauge invariant} quantity. This fixes $\vec{\mathbf{e}}^{\perp} = - \dot{\vec{\mathbf{a}}}^{\perp}$ for any gauge choice. In the Lorentz gauge, $\vec{\mathbf{a}}=\vec{\mathbf{a}}^{\parallel}+ \vec{\mathbf{a}}^{\perp}$ ($\nabla \cdot \vec{\mathbf{a}} \neq 0$), hence the vector potential does contribute a term $- \dot{\vec{\mathbf{a}}}^{\parallel}$ to $\vec{\mathbf{e}}^{\parallel}$ (Eq. \ref{scalar_phi}).

\section*{Problems}
\addcontentsline{toc}{section}{Problems}
%

\begin{prob}(*)
\label{prob8_p1_ce}
Show that such the choice in Eq. \ref{Lorentz_gauge} can always be made.
\end{prob}

\begin{prob}(*)
\label{prob9_p1_ce}
Show that such the choice in Eq. \ref{Coulomb_gauge} can always be made. 
\end{prob}








\backmatter

\Extrachap{Solutions}

\section*{Problems of Chapter~\ref{p1_ce}}

\begin{sol}{prob1_p1_ce}
First, convert [Oe] $\rightarrow$ [A]/[m].
$$\mid \vec{\mathbf{H}}_0\mid = 1900 ~ {\rm [Oe]} = 1900 \times {10^{3}\over 4 \pi} {\rm [A]/[m]}.$$
Then the magnitude of the magnetic flux density is:
$$\mid \vec{\mathbf{B}}_0\mid= \mu_0 \left ( {{\rm [T]}\over {\rm [A]/[m]}} \right ) \cdot \mid \vec{\mathbf{H}}_0\mid {\rm [A]/[m]}$$
$$\Longrightarrow \mid \vec{\mathbf{B}}_0\mid = 4 \pi \times 10^{-7} \times 1900 \times {10^{3}\over 4 \pi} ~ {\rm [T]}$$
$$\Longrightarrow \mid \vec{\mathbf{B}}_0\mid=  0.19 {\rm [T]}$$

Notice, therefore, that you can easily ``convert'' from magnetic field intensity in [Oe] 
to magnetic flux density in [T] by multiplying the former by the factor $10^{-4}$.
\end{sol}

\begin{sol}{prob2_p2_ce}
\noindent{\bf a)} We fix the direction of the electric field and the initial velocity of the electron along the $x$-axis. The initial position of the electron is set at the origin of the $x$-axis. The motion is unidimensional, given by:

\begin{equation}
m {\ud ^2 x \over \ud t^2} = -e E_x,
\end{equation}

\noindent where $m$ is the mass of the electron, and $e$ is the fundamental charge. Notice that we write explicitly the negative sign, as the electron charge is $e^{-}\equiv-e$. We integrate twice and obtain at each step:

\begin{equation}
v_x \equiv  {\ud x \over \ud t}= - {e \over m} E_x t + c_1;
\end{equation}
\begin{equation}
x = - {1 \over2} {e \over m} E_x t^2 + c_1 t + c_2,
\end{equation}

\noindent where $c_1$ and $c_2$ are constants of integration. As we assume that the electron is at the origin of coordinates $x=0$ at the initial time $t_i$, where it has velocity $v_i$, we may insert these values in the equations above, finding:

\begin{equation}
c_1 = v_i +  {e \over m} E_x t_i,
\end{equation}
\begin{equation}
c_2 = -v_i t_i - {1 \over2} {e \over m} E_x t^2_i.
\end{equation}

Hence, we obtain the velocity and position solutions:

\begin{equation}
v_x= - {e \over m} E_x (t-t_i) + v_i; \label{v_e_eqn}
\end{equation}
\begin{equation}
x = - {1 \over2} {e \over m} E_x (t-t_i)^2 + v_i(t-t_i). \label{x_e_eqn}
\end{equation}

There are a few issues to notice. First, fixing $v_i>0$ and $E_x>0$, the solutions found are formally analogous to that of a free-falling particle in a gravitational field\footnote{In the case of the gravitational field, near the ground level, we take $\vec{\mathbf{g}} = -g \hat{\mathbf{x}}$, giving $$x = -1/2 g (t-t_i)^2+ v_i(t-t_i).$$}. The electron, in this case, has the trajectory of a projectile. However, for $v_i>0$ and $E_x<0$, the electron's velocity increases monotonically with time.

\noindent {\bf b)} Let us set $t_i \equiv 0$ without loss of generality. We will consider the electron mass as approximately $m \sim 9.11 \times 10^{-31}$ [kg] and the fundamental charge as $e \sim 1.60 \times 10^{-19}$ [C]. Then the ratio $m/e \sim 5.68 \times 10^{-12}$ [kg]/[C]. Notice then, by Eq. \ref{v_e_eqn},  that the conditions of the problem require  (when $v=0$):
$$v_i \mid_{v=0} = {e \over m} E_x t \mid_{v=0}.$$
\noindent Therefore, we may set $E_x$ in units of the ratio $m/e$ per unit of time. Moreover, since we require that
$$\alpha  =  {e \over m} E_x \alpha \Longrightarrow E_x = {m \over e},$$

\noindent that is,  the value of the electric field intensity should be of the same as the ratio ${m \over e}$. Notice that $E_x$  is given in units of [N]/[C] = [kg][m][s]$^{-2}$/[C], then we may set a normalization unit [$\mathcal{E}$] = $5.68 \times 10^{-12}$ [kg][m][s]$^{-1}$/[C], so that the electric field intentisty in this new unit is given by $E_x$ =  1 [$\mathcal{E}$]/[s], also meeting the requirements of the problem.

\noindent {\bf c)} From Eq. \ref{x_e_eqn} and the conditions set for item {\bf b}, we have, for $E_x$ in units of [$\mathcal{E}$]/[s]:

\begin{equation}
x = - {1 \over2} t^2 + v_i t, E_x = 1 ;
\end{equation}
\begin{equation}
x = +{1 \over2} t^2 + v_i t,  E_x = -1;
\end{equation}

\noindent {\bf d)} Setting $x=l$, $t_i=0$, $v_i=0$ in Eq. \ref{x_e_eqn}, we find that:
\begin{equation}
l = - {1 \over2} {e \over m} E_x t^2,
\end{equation}
\begin{equation}
t= \sqrt{-2 l m \over E_x e}.
\end{equation}

\noindent So we eliminate the time variable by inserting the latter equation into Eq.  \ref{v_e_eqn}:
\begin{equation}
v= {e \over m} (-E_x)   \sqrt{-2 l m \over  E_x  e}. 
\end{equation}
\noindent Assuming $E_x<0$ we find:
\begin{equation}
v= \sqrt{2 e |E_x| l \over m}, E_x<0, v_i=0.
\end{equation}

\noindent Notice that for the same previous conditions ($t_i=0$, $v_i=0$), but $E_x>0$, we again have the same result for the velocity, but with opposite sign. Notice also that the square root will {\sl not} be negative in this case because the electron will only attain negative position $x=-l$ values (we are always fixing the initial position of the electron at $x=0$). Then we have in this case:

\begin{equation}
v= - \sqrt{2 e |E_x| |l| \over m}, E_x>0, v_i=0.
\end{equation}
\end{sol}

\begin{sol}{prob3_p1_ce}
Let a charge $q_1$ be located at the origin of a coordinate system. Given the spherical symmetry the charge distribution (which is trivial in this case, as it is just one charge), we choose a spherical coordinate system. The corresponding electric field should depend on the radial coordinate $r$, and should not depend of the angular coordinates $\theta$ and $\phi$. Let us evaluate for the present case the surface integral at an arbitrary radius, as stated by Gauss's integral law (Eq. \ref{Gauss_int_eqn}):

\begin{equation}
\oint_{S} \epsilon_0 \vec{\mathbf{E}}_0 \cdot \ud \vec{\mathbf{S}} = 
\int_0^{\pi} \left [
\int_0^{2\pi} \epsilon_0 E_r (r \sin \theta \ud \phi) 
\right ]
 (r \ud \theta) = \epsilon_0 E_r 4\pi r^2.
\end{equation}
As the whole charge distribution is centered at the origin, the volume integral [Eq. \ref{Gauss_int_eqn}] simply furnishes the charge $q_1$. Therefore,

\begin{equation}
\epsilon_0 E_r 4\pi r^2 = q_1 \Longrightarrow \vec{\mathbf{E}}_0^1 = {q_1 \over  4\pi\epsilon_0 r^2}
\hat{\mathbf{r}}. \label{Epontual}
\end{equation}
According to the Lorentz force law (Eq. \ref{Lorentz_eqn}), at the location of $q_2$, the electric field intensity $\vec{\mathbf{E}}_0^1$ generated by charge $q_1$ introduces a force on $q_2$ given by:

\begin{equation}
\vec{\mathbf{F}}_{12} = q_2 \vec{\mathbf{E}}_0^1 = q_2 \left ( {q_1 \over  4\pi\epsilon_0 r^2} \hat{\mathbf{r}} \right ).
\end{equation}
That is Couloumb's law.
\end{sol}

\begin{sol}{prob4_p1_ce}
If we apply Amp\`ere's law for a {\sl closed} surface, it means that the contour will close on itself, in the limit of which it tends to a point, that is, the contour integral tends to zero and the open surface integrals become closed surface integrals:
\begin{equation}
\oint_{S} \vec{\mathbf{J}} \cdot \ud \vec{\mathbf{S}} + {\ud \over \ud t} \oint_{A} \epsilon_0
\vec{\mathbf{E}}_0 \cdot \ud \vec{\mathbf{S}} = 0.
\end{equation}
But from Eq. \ref{Gauss_int_eqn}, the second term integral above is given by the volume integral of the charge, and therefore we prove the continuity equation (Eq. \ref{Cont_int_eqn}).
\end{sol}

\begin{sol}{prob5_p1_ce}
From Faraday's law (Eq. \ref{Faraday_int_eqn}), a region where the electric field intensity does not have a circulation implies
\begin{equation}
\oint_{C}\vec{\mathbf{E}}_0\cdot \ud \vec{\mathbf{l}} = 0,
\label{E_nocirc}
\end{equation}
which means that 
\begin{equation}
{\ud \over \ud t} \int_{S} \mu_0 \vec{\mathbf{H}}_0 \cdot \ud \vec{\mathbf{A}} = 0,
\end{equation}
that is, the time rate of change of the magnetic flux density is neglectable. This condition is met in {\sl quasi-electrostactic systems}. 

An interesting aspect of that situation is the fact that, whatever the contour adopted, 
Eq. \ref{E_nocirc} is always zero, which means that the path integral betweeen two arbitrary points does not depend on the chosen path. The {\sl electromotive force} is defined as the path integral between two points $a$ and $b$:
\begin{equation}
\epsilon_{ab} = \int_a^b \vec{\mathbf{E}}_0 \cdot \ud \vec{\mathbf{l}}. \label{eletromotriz}
\end{equation}
For a region where the electric field intensity does not have a circulation, the calculation of the electromotive force between two points, therefore, does not depend on the chosen path between those points. In that case, the electromotive force is called the {\sl voltage} between two points.

An example of a region where the electric field intensity does not have a circulation is that between two parallel sheets with uniform charge density, generating a static electric field intensity between them.
\end{sol}

\begin{sol}{prob8_p1_ce}
Suppose the given potentials did not obey Eq. \ref{Lorentz_gauge}. Then we could adjust them through the gauge transformations (Eq. \ref{gauge_trans}) so that they do:

\begin{equation}
\nabla \cdot \vec{\mathbf{a}}^{\prime} =
- {1 \over c^2} {\partial \over \partial t}
\left ( \phi^{\prime} \right )
\Rightarrow
\nabla \cdot \left ( \vec{\mathbf{a}} + \nabla \chi \right )  = 
- {1 \over c^2} {\partial \over \partial t}
\left ( \phi - {\partial \chi \over \partial t } \right )
\end{equation}

\begin{equation}
\Longrightarrow ~ \left ( \nabla^2 - {1 \over c^2} {\partial^2 \over \partial t^2} \right ) \chi  = - \nabla \cdot  \vec{\mathbf{a}}- {1 \over c^2} {\partial \phi \over \partial t},
\end{equation}

\noindent which means that, by finding a solution to the equation above for gauge function $\chi$, one can find the appropriate pair of potentials $(\vec{\mathbf{a}}^{\prime},\phi^{\prime})$ satisfying Eq. \ref{Lorentz_gauge}. 
\end{sol}

\begin{sol}{prob9_p1_ce}
Suppose first $\nabla \cdot \vec{\mathbf{a}} \neq 0$. Then transform to $\vec{\mathbf{a}}^{\prime} = \vec{\mathbf{a}} + \nabla \chi$, where we desire $\nabla \cdot \vec{\mathbf{a}}^{\prime} = 0$. That is: $\nabla \cdot (\vec{\mathbf{a}} + \nabla \chi) = 0 \rightarrow \nabla^2 \chi = - \nabla \cdot \vec{\mathbf{a}}$, which is a Poisson equation for $\chi$. Once solved, one has a solenoidal $\vec{\mathbf{a}}^{\prime}$.
\end{sol}


\bibliographystyle{plain}
\bibliography{CBSQED}


\end{document}